\documentclass[aps,prd,notitlepage,twocolumn,reprint,superscriptaddress,nofootinbib,longbibliography]{revtex4-1}
\usepackage{multirow}
\usepackage{amssymb, esvect, amsmath, graphicx, latexsym, amsthm, slashed, eso-pic}
\usepackage{xcolor}
\usepackage{amsmath,amsthm,amsfonts,amscd,amssymb} 
\usepackage{eucal}
\usepackage{braket}
\usepackage{slashed}
\usepackage{hyperref}
\usepackage{graphicx}
\usepackage{bm}
\usepackage{siunitx}
\usepackage{xcolor}
\usepackage{afterpage}
\usepackage{changepage}
\usepackage{sidecap}

\usepackage{hyperref}
   \DeclareMathOperator{\gev}{GeV}    \DeclareMathOperator{\cm}{cm}

   \def\oL{\overline} 
       
\newcommand{\beq}{\begin{equation}} \newcommand{\eeq}{\end{equation}}
\newcommand{\bea}{\begin{eqnarray}} \newcommand{\eea}{\end{eqnarray}}

\def\lsim{\mathrel{\raise.3ex\hbox{$<$\kern-.75em\lower1ex\hbox{$\sim$}}}}
\def\gsim{\mathrel{\raise.3ex\hbox{$>$\kern-.75em\lower1ex\hbox{$\sim$}}}}

\newcommand{\be}{\begin{eqnarray}}
\newcommand{\ee}{\end{eqnarray}}

\newcommand{\benum}{\begin{enumerate}}
\newcommand{\eenum}{\end{enumerate}}
\newcommand{\bi}{\begin{itemize}}
\newcommand{\ei}{\end{itemize}}

\begin{document}

\preprint{FERMILAB-PUB-19-082-AE-T}

\title{Gravitational Direct Detection of Dark Matter}

\author{Daniel Carney}\thanks{E-mail: carney@umd.edu}
\affiliation{Joint Quantum Institute/Joint Center for Quantum Information and Computer Science, \\ University of Maryland, College Park/National Institute of Standards and Technology, Gaithersburg, MD, USA}
\affiliation{Fermi National Accelerator Laboratory, Batavia, IL, USA}
\author{Sohitri Ghosh}
\affiliation{Joint Quantum Institute/Joint Center for Quantum Information and Computer Science, \\ University of Maryland, College Park/National Institute of Standards and Technology, Gaithersburg, MD, USA}\author{Gordan Krnjaic}
\affiliation{Fermi National Accelerator Laboratory, Batavia, IL, USA}
\author{Jacob M. Taylor}\thanks{E-mail: jmtaylor@umd.edu}
\affiliation{Joint Quantum Institute/Joint Center for Quantum Information and Computer Science, \\ University of Maryland, College Park/National Institute of Standards and Technology, Gaithersburg, MD, USA}

\medskip

\date{\today}

\begin{abstract}
The only coupling dark matter is guaranteed to have with the standard model is through gravity. Here we propose a concept for direct dark matter detection using only this gravitational coupling. We suggest that an array of quantum-limited mechanical impulse sensors may be capable of detecting the correlated gravitational force created by a passing dark matter particle. We consider the effects of irreducible noise from couplings of the sensors to the environment and noise due to the quantum measurement process. We show that the signal from Planck-scale dark matter is in principle detectable using a large number of gram-scale sensors in a meter-scale array with sufficiently low quantum noise, and discuss some experimental challenges en route to achieving this target.
\end{abstract}

\maketitle

\tableofcontents

\section{Introduction}

Observations of galactic rotation curves \cite{Sofue:2000jx}, gravitational lensing \cite{Massey:2010hh}, the cosmic microwave background \cite{Aghanim:2018eyx}, galaxy cluster collisions \cite{Markevitch:2003at}, and the large scale structure of our universe \cite{Primack:2015kpa} are  inconsistent with a model of the universe containing only general relativity and the standard model of particle physics. Positing the existence of cold dark matter (DM) successfully explains these diverse, independent observations (see \cite{Bertone:2016nfn} for a review). However, despite decades of dedicated searches, dark matter has yet to be directly detected in the neighborhood of Earth \cite{Tanabashi:2018oca}.
  
Existing approaches to direct detection are insensitive to DM scattering via the gravitational force. Instead, these test the additional hypothesis that DM interacts with visible matter through much stronger non-gravitational forces, and that DM is in a range of relatively light masses. Here we propose a new direct detection technique based entirely on DM's required gravitational interactions, completely independent of any non-gravitational couplings. If this target can be achieved, it would open an entirely new mode of DM search in a largely unexplored mass range.

Our proposed strategy is to build a three-dimensional array of force sensors. A heavy DM particle passing through the array will exert a small but \emph{correlated} force on the sensors nearest its trajectory. Much like tracking a particle in a bubble chamber, we can then pick out this correlated force signal along the DM ``track'' through the array. In particular, this means that the detector gains complete directional information, allowing for robust rejection of many traditional DM detection backgrounds. 

Since the gravitational interaction strength increases linearly with mass of both the passing DM and the sensor, we suggest the use of macroscopic mechanical force sensing devices. Driven in large part by LIGO \cite{TheLIGOScientific:2014jea}, the last few decades have witnessed dramatic improvements in the continuous quantum-limited sensing of mechanical systems with masses ranging from single ions to tens of kilograms. Numerous devices have been demonstrated with sensing at the ``standard quantum limit'' (SQL) \cite{caves1981quantum}, and advanced techniques such as the use of squeezed light \cite{purdy2013strong,clark2017sideband,aasi2013enhanced} or backaction evasion \cite{braginsky1980quantum,BRAGINSKY1990251,clerk2008back,hertzberg2010back,khosla2017quantum,danilishin2018new} have achieved even lower noise levels.

The sensitivity of the array is set by various sources of noise acting on the devices, and the core goal of this paper is to study these noise limitations. In particular, we focus on two key, irreducible noise sources: coupling of the sensors to their thermal environment, and quantum measurement noise coming from the Heisenberg principle. While numerous additional technical noise sources--stray fields, laser instabilities, collective modes of an array, and so forth--are inevitable, these are ultimately avoidable by sufficiently clever experimentalists. On the other hand, thermal and quantum noise set the fundamental floor for any experiment. Indeed, many experiments, for example advanced LIGO, now operate in a regime in which quantum measurement noise is the dominant factor setting their sensitivity. Here we study the extent to which this fundamental noise floor would allow for gravitational direct detection of dark matter.

\begin{figure}
\includegraphics[scale=.65]{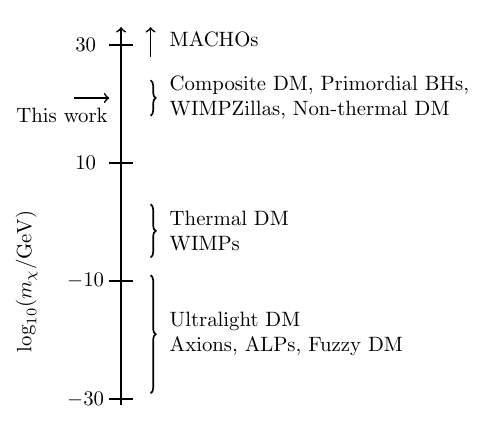}
\caption{Broad classification of viable dark matter models according to mass. In this work, we focus on dark matter candidates around $m_{pl} \sim 10^{19} ~ \gev$ and above.}
\label{scales}
\end{figure}

In order to maximize the signal from the passing dark matter, we suggest an array of gram-scale devices spaced at centimeter-scale distances. At these scales, we find that with well-isolated systems (eg. at high vacuum and/or dilution refrigeration temperatures), quantum noise at the SQL dominates over thermal noise. This calls for the use of advanced measurement techniques which can evade the SQL, as we detail below. However, assuming the use of a sufficiently noise-evading measurement protocol, we find that an array of around $10^8-10^9$ sensors could in principle detect any DM candidate with mass heavier than around $m_{\rm Pl} \sim 10^{19} \gev$.

There are many viable models of DM in our window of detectability. Some examples include WIMPzillas \cite{Kolb:1998ki,Kolb:2017jvz,Kolb:2007vd,Harigaya:2016vda}, GUT-scale coannihilating particles \cite{Berlin:2017ife}, Planckian interacting DM \cite{Garny:2015sjg}, composite ``nuclear" DM with large occupation numbers \cite{Krnjaic:2014xza,Hardy:2014mqa,Hardy:2015boa,Redi:2018muu}, dark quark nuggets \cite{Detmold:2014qqa,Detmold:2014kba,Gresham:2017zqi,Gresham:2017cvl,Gresham:2018anj,Bai:2018dxf}, Planckian relics from evaporated black holes \cite{aharonov1987unitarity,MacGibbon:1987my} or even small extremal black holes \cite{Bai:2019zcd}. While there are some recent proposals to detect the non-gravitational interactions of specific ultraheavy DM candidates \cite{Grabowska:2018lnd,Bramante:2018tos,Bramante:2018qbc}, most viable DM candidates in this mass range have extremely feeble non-gravitational interactions with visible matter. A mature realization of our concept can robustly test all such models without invoking any non-gravitational interaction. 

We note two recent works looking to detect DM gravitationally \cite{hall2018laser,kawasaki2019search}. Both suggest using a single sensor without a noise-evading measurement protocol, leading to comparatively limited detection reach and lack of background event rejection.

\section{Detector paradigm}

\begin{figure*}[t]
 \includegraphics[scale=0.7]{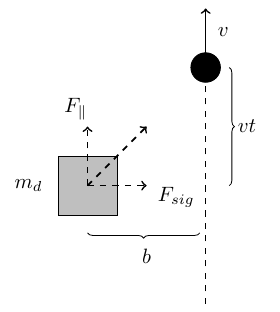}  ~ \includegraphics[scale=0.2]{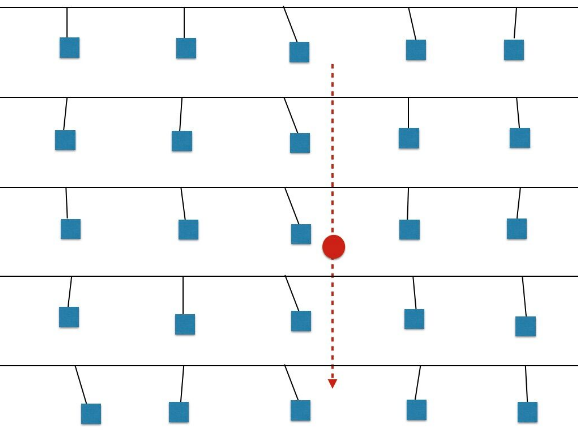} ~~ \includegraphics[scale=0.27]{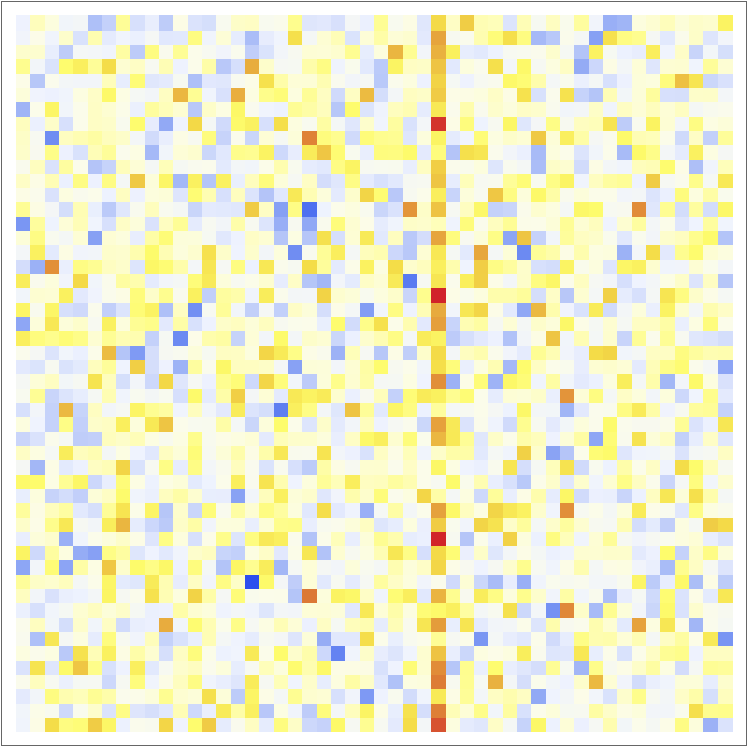} ~~~ \includegraphics[scale=.40]{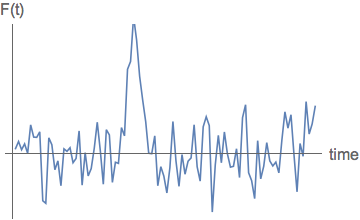}
  \caption{Left: Kinematics of the DM-sensor scattering event, viewed from above the scattering plane. Center left: Schematic of (a cross-section of) the detector array, with suspended pendula used as mechanical resonators. As the DM passes through the array, it produces a correlated impulse on the sensors nearest its track. This diagram suppresses the readout mechanism (see figure \ref{velocitycartoon}), for which there are many potential implementations. Center right: Simulation of an event on a $50 \times 50$ plane of sensors. The colors represent impulses; blue are impulses to the left while red are to the right. The track of yellow-red corresponds to the signal. Right: Cartoon of single-sensor data stream, with an event.}
\label{cartoon}
\end{figure*}

We begin by giving an overview of the detector concept and methods for estimating its sensitivity to DM candidates of various masses. The primary goal is to derive the requirements on the detector such that it has non-trivial detection reach to sufficiently heavy dark matter. We will see that achieving this goal will require overcoming a number of technical challenges, which are discussed in detail in section \ref{section-challenges}.

The essential idea is to continuously monitor a three-dimensional array of mechanical sensors. Each sensor has its position (or momentum) read out continuously, as done for example in LIGO. When a heavy object passes through the array, for example a heavy dark matter particle, it will exert a small gravitational force on each sensor, causing slight deviation in the sensor motions. The sensors which are nearest this passing object will have the largest deviation, forming a ``track'' through the array. See figure \ref{cartoon} for some visualizations of this process.

Before discussing the array, we begin by studying the interaction of a passing DM particle with a single sensor. See figure \ref{cartoon} for a diagram of the kinematics. We are interested in the Newton force $\mathbf{F}_N = G_N m_{\chi} m_{s} \hat{\mathbf{r}}/r^2$ between a sensor of mass $m_{s}$ and DM particle of mass $m_{\chi}$. A lab at rest on Earth sees the DM pass by with average ``wind speed'' $v \approx 220~\si{\kilo\meter}/\si{\second}$. Thus the DM imparts momentum to the detector on a very short timescale $\tau$. For a representative impact parameter $b$ of approximately a millimeter, we have $\tau \approx b/v \sim 10^{-8}~\si{\second}$. 

The fundamental limitation to sensing this tiny Newtonian force is the presence of other, noisy forces acting on the sensor. The total force on the sensor is
\be
F_{\rm in}(t) = F_{\rm sig}(t) + F_{\rm T}(t) + F_{\rm M}(t).
\ee
The noise terms $F_{\rm noise} = F_{\rm T} + F_{\rm M}$ are random variables. The measurement-added noise $F_{\rm M}$ is a fundamental quantum limitation, and depends on the system observable we probe and how precisely we perform the readout (see section \ref{section-sensitivities} for details). Meanwhile, the thermal noise $F_{T}$ is set by the ambient temperature $T$ and the nature of the thermal bath coupling to the detectors, but independent of the measurement readout scheme. 

In our continuous monitoring protocol, the data comes in the form of a timeseries $F(t)$ for each sensor. For the signal, we will take as our basic observable the total impulse delivered to the sensor along the axis transverse to the dark matter trajectory (see figure \ref{cartoon}),
\be
\label{signal}
F_{\rm sig} = \frac{G_N m_{\chi} m_{s} b}{(b^2 + v^2 t^2)^{3/2}}.
\ee
The total impulse is easy to calculate:
\be
I = \int_{-t_{\rm int}/2}^{t_{\rm int}/2} dt \ F_{\rm sig}(t) \to 2 G_N m_{\chi} m_s \tau/b^2 = \overline{F} \tau,
\ee
where $t_{\rm int}$ is an integration window, $\overline{F}$ is the average force, and we assume $t_{\rm int} \gtrsim \tau$ and sufficient incoming velocity so that we can approximate the DM as moving on a straight-line track. Here we have chosen the transverse component because its time integral is finite, but one could filter the data $F(t)$ with an appropriate function and use instead the time-integral of this filtered data. In this sense one can look for any particular component of the force. See section \ref{section-data} for details on data processing issues. 

Note that by the equivalence principle, only tidal forces are observable. To use equation \eqref{signal}, it is critical that the readout be referenced to a system sufficiently far from the sensor such that the Newtonian acceleration produced on the reference is negligible.

To estimate our sensitivity to the signal \eqref{signal}, we need to compare its size to the noise acting on the sensors. The noise is characterized by the variance $\braket{\Delta I^2} = \int \int dt dt' \ \braket{F_{\rm noise}(t) F_{\rm noise}(t')}$. For stationary noise, this correlation function is proportional to $\delta(t-t')$. Thus the RMS impulse grows as a square root in time
\be
\label{measnoise}
\Delta I^2 = \alpha t_{\rm int},
\ee
for some constant $\alpha$. This time dependence is characteristic of Brownian motion. Since the integrated signal strength grows approximately linearly in $t_{\rm int}$ while the DM is nearby and the noise only grows as $\sqrt{t_{\rm int}}$, an appropriately chosen $t_{\rm int} \approx \tau$ serves to average out the fluctuations caused by the noise.

The signal-to-noise ratio of a single sensor during a passing DM event is therefore ${\rm SNR}^2 = I^2/\Delta I^2 = \oL{F}^2 \tau/\alpha$. Now consider constructing an array with $N_{\rm det}$ of these sensors. As the DM passes through the array, it will pass by a ``track'' of $N \sim N_{\rm det}^{1/3}$ of the sensors. For sensors spaced far enough from each other, thermal noise is uncorrelated amongst the sensors, and we further assume that we separately monitor each sensor so that measurement-added noise is also uncorrelated. In practice there will also be some correlated sources of noise; here we assume this is sub-dominant to the single-sensor noise, and refer the reader to section \ref{section-correlations} for a more extensive discussion. We then ask the statistical question: given a fixed track of $N$ sensors, did they all receive enough impulse to be seen above their individual noises? This is answered by adding the SNR of each sensor in quadrature, so the standard error decreases like $1/\sqrt{N}$, and the signal-to-noise ratio is given by
\be
\label{SNRbasic}
\text{SNR}^2 = N \overline{F}^2 \tau/\alpha.
\ee
It is critical that the signal here is the entire, correlated track of moving sensors. This in particular means that our backgrounds--that is, events other than passing DM which likewise trigger a correlated track of displacements--are very different from traditional direct detection experiments (see section \ref{section-backgrounds} for more details on background rejection). It also means that the signal includes complete directional information.

\begin{figure}
\includegraphics[scale=0.37]{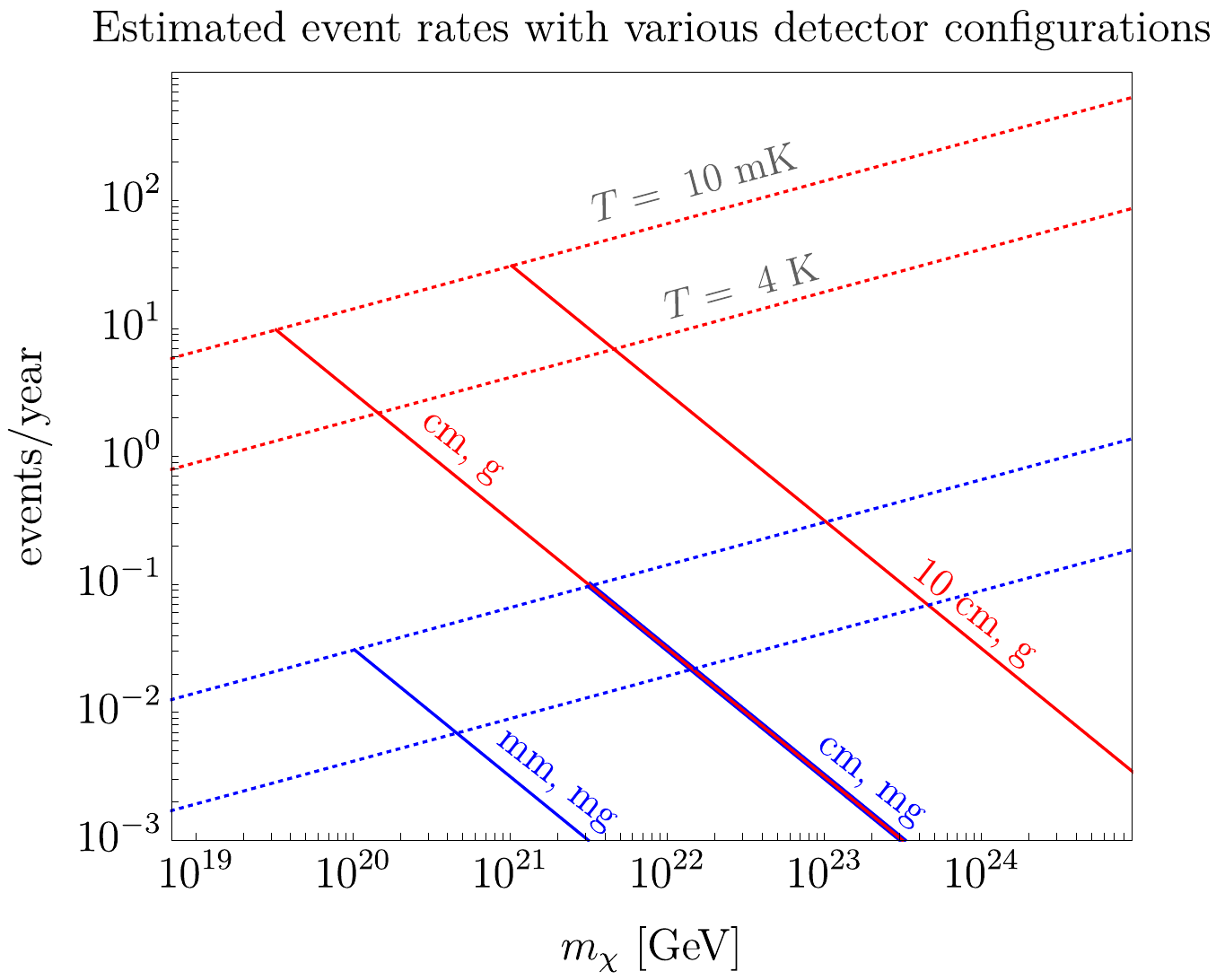}
\caption{Detectable DM event rates, with a variety of detector configurations. Thick lines correspond to number of events per year, assuming all DM particles have mass $m_{\chi}$. The 1/$m_{\chi}$ falloff in the rate is due purely to number flux (see equation \eqref{rates}); by construction, all DM candidates passing through the detector are detected with $5\sigma$ confidence. Solid lines are labeled by the array lattice spacing (mm, cm, or $10$ cm) of the detector and individual sensor masses (milligram in blue or gram in red).   Dashed lines labeled by temperature ($4$ K or $10$ mK) demonstrate the increased sensitivity of our scheme with improved environmental isolation. Here we are assuming background gas-limited environmental noise with the same fiducial parameters as in \eqref{SNRfinal-falling}.}
\label{plots} 
\end{figure}

Our basic result \eqref{SNRbasic} can be used to estimate the SNR for any particular detector scheme. Given a target DM mass, the SNR is set entirely by the noise on our detector. Thermal coupling of the sensors to their environment sets the ultimate, irreducible technical noise floor--one can environmentally isolate the system as much as possible, but never completely. On the other hand, measurement-added noise can be reduced by a variety of techniques, as discussed below. Thus, to understand the fundamental limits on gravitational DM detection, we begin by estimating our SNR under the assumption that we have sufficiently reduced measurement-added noise to be subdominant to the thermal component. It is important to understand that this is an ambitious experimental target--see section \ref{section-sensitivities} for an extensive discussion on challenges to reaching the thermal floor in macroscopic sensors.

For detectors mechanically coupled to a support structure at temperature $T$, we have $\alpha_{\rm mech} = 4 m_s k_B T \gamma$ with $\gamma$ the detector's mechanical damping rate \cite{clerk2010introduction}. For freely-falling detectors, we are limited instead by the latent gas pressure $P$, which gives $\alpha_{\rm gas} = P A_d \sqrt{m_a k_B T}$, where $A_d$ is the cross-sectional area of each detector and $m_a$ is the mass of the gas atoms \cite{green1951brownian}. Numerically, we thus obtain the following estimates for the \text{SNR}:
\begin{align}
\begin{split}
\label{SNRfinal-mech}
\text{SNR}^2 &= \frac{G_N^2 m_{\chi}^2}{v} \frac{L}{d^4} \frac{m_s}{k_B T \gamma} \\
&\approx  10 \times \left( \frac{m_{\chi}}{1~\si{\milli\gram}} \right)^2 \left( \frac{m_s}{1~\si{\milli\gram}} \right) \left( \frac{1~\si{\milli\meter}}{d} \right)^4
\end{split}
\end{align}
in the case of detectors mechanically coupled to a support structure, and 
\begin{align}
\begin{split}
\label{SNRfinal-falling}
\text{SNR}^2 & = \frac{4 G_N^2 m_{\chi}^2}{v} \frac{L}{d^4} \frac{m_s^2}{P A_d \sqrt{m_a k_B T}} \\
& \approx 10^4 \times \left( \frac{m_{\chi}}{1~\si{\milli\gram}} \right)^2 \left( \frac{m_s}{1~\si{\milli\gram}} \right)^2 \left( \frac{1~\si{\milli\meter}}{d} \right)^4,
\end{split}
\end{align}
for freely-falling detectors. Here for simplicity we assumed a cubical array of side length $L$ (so that the number of sensors nearest the DM path is $N \sim L/d$ and the total number of sensors $N_{tot} = (L/d)^3$) with $L = 1~\si{\meter}$, and assumed dilution fridge temperatures $T = 10~\si{\milli\kelvin}$, helium ion-pump vacuum pressures $P = 10^{-10}~\si{\pascal}, m_a = 4~\si{\amu}$ \cite{thompson1977characteristics}, mechanical damping $\gamma = 10^{-8}~\si{\hertz}$ \cite{adams1991use}, and typical solid density $\rho_{\rm s}\sim 10~\si{\gram}/\si{\centi\meter}^3$ for the detectors.

The signal-to-noise ratios \eqref{SNRfinal-mech}, \eqref{SNRfinal-falling} represent our fundamental detection sensitivities. Crucially, the detection is deterministic: if a sufficiently heavy DM particle passes through the detector, and we demand our detector parameters are such that ${\rm SNR} > 5$, it will \emph{always} will be detected with $5\sigma$ confidence. The number of DM events we have per year is then entirely determined by the number density of the DM. The observed local DM density $\rho_{\chi} \approx 0.3~\gev/\cm^3$ \cite{Bovy:2012tw} means that, for a detector array of total cross-section $A_d$, the rate of DM passing through the detector is
\be
\label{rates}
R = \frac{ \rho_{\chi} v A_d}{m_\chi} \sim \frac{1}{\rm year} \left( \frac{m_{\rm Pl}}{m_{\chi}} \right) \left( \frac{A_d}{1 \, \rm m^2} \right).
\ee
In figure \ref{plots}, we plot our predicted event rates with a variety of detector geometries, with $10^9$ detectors. With a billion detectors at the gram scale, Planck-scale gravitational DM detection is achievable.  Reaching heavier masses can be achieved with a sparser array.

The above estimates should be interpreted as a long-term target subject to further possible developments. There are a number of technical challenges which need to be overcome to realize these estimates, which we discuss in detail in section \ref{section-challenges}. Our central message is really that the rules of measurement in quantum mechanics allow for the required sensitivities: it is not inconceivable that one could build an appropriate apparatus and perform gravitational direct detection searches of heavy dark matter. 

We note also that there are numerous ways one can imagine improving the situation from that considered here. Advanced measurement techniques involving coherent readout \cite{micadei2015coherent} or error correction \cite{kessler2014quantum,zhou2018achieving} can significantly improve the detection sensitivity. One can also relax the need for $5\sigma$ detection of each individual track event and look for statistics to build up over a long time (say, a few years of exposure) for the evidence of tracks, analagous to statistical evidence for WIMP events in a heavy noble detector. Pursuing these types of techniques is a subject of active work, beyond the scope of this introductory paper.

\section{Implications for the DM landscape}
Before moving on to discuss technical issues in the experimental realization of these ideas, we make some brief comments on the implications for such an experiment in the broader search for dark matter. As emphasized above, the scheme relies only on the gravitational coupling of DM to visible matter, so if the required sensitivity can be achieved, the experiment would either discover or rule out \emph{any dark matter candidate in the appropriate range of mass}.

Our detector concept is capable of searching for DM candidates around and above the Planck mass. At this scale, DM is presumably not a fundamental particle. Viable options include composite objects like dark nuclei or dark quark nuggets \cite{Krnjaic:2014xza,Hardy:2014mqa,Hardy:2015boa,Gresham:2017zqi,Redi:2018muu,Detmold:2014qqa,Detmold:2014kba,Gresham:2017zqi,Gresham:2017cvl,Gresham:2018anj,Bai:2018dxf}, extended objects like topological defects, or quantum gravity exotica like primordial black hole remnants \cite{aharonov1987unitarity,MacGibbon:1987my,Bai:2019zcd}. In the most optimistic scenario, a large-scale version of our proposal could reach down to feebly-interacting GUT-scale DM candidates \cite{Kolb:1998ki,Kolb:2017jvz,Kolb:2007vd,Harigaya:2016vda,Berlin:2017ife}, which could be fundamental particles.

A confirmed signal would then imply a rich cosmological history for this sector, which must contain either a modified inflationary potential, an early phase of DM self-assembly, a non-thermal abundance generation mechanism, or a dark-sector phase transition. At least one of these mechanisms must be active to realize masses for the DM in these scenarios. Conversely, a null result would have far reaching consequences for this class of models by excluding a broad swath of DM candidate masses, independently of their other non-gravitational interactions.

\section{Technical challenges}
\label{section-challenges}

A detailed realization of the detector concept outlined above will necessarily come with a large set of technical challenges to be overcome. Many of these would depend sensitively on the chosen sensor platform--microwave versus optical domain, sensors which are superconducting or not, and so forth. Here, we discuss a number of further issues which would be largely independent of any particular experimental choices, again focusing on the fundamental limits to achieving the desired sensitivity. We discuss backgrounds in the sense of a traditional particle physics detection experiment, requirements for achieving the necessary reduction in quantum measurement noise, correlations between sensors, and make some preliminary remarks about data processing and filtering issues. We end with a few comments on possible choices of concrete sensor architectures.

\subsection{Backgrounds}
\label{section-backgrounds}

In direct DM detection experiments, background events pose a serious problem and need to be systematically understood. For example, in a xenon-based detector, the signal is that a DM particle scatters elastically off a xenon nucleus, causing the xenon atom to recoil and emit a photon. The photon is the signal. The issue is that any number of other events, having nothing to do with DM, could cause the same signal.

Unlike traditional direct DM detection experiments, the signal in our proposal does not consist of a single recoiling object, but rather a correlated track worth of displaced, macroscopic objects. This allows for rejection of many typical backgrounds. For example, cosmic rays only hit an individual sensor. More challenging issues come from correlated backgrounds like seismic noise or propagating signals induced by sensor-sensor interactions. The latter should be suppressed at the lattice spacings used here; more generally, these types of backgrounds should have vastly different propagation speed from our DM signal $v \sim 220 ~ {\rm km/s}$. While a detailed study of backgrounds will be needed for a mature experimental realization, the basic characteristics of the signal in our proposal offer an extremely promising route to robust background rejection.

\subsection{Achieving thermally-limited sensitivities}
\label{section-sensitivities}

Our optimal measurement sensitivities \eqref{SNRfinal-mech}, \eqref{SNRfinal-falling} were derived assuming that thermal noise dominates over measurement-added noise. We now turn to an analysis of the feasibility of achieving sufficiently low measurement noise for this approximation to be accurate. For an introduction to the topic of quantum measurement noise, we refer the reader to the excellent review \cite{clerk2010introduction}. For a self-contained treatment of noise in a linear optomechanical device, see the appendix of \cite{Ghosh:2019rsc}.

In the continuous measurement of a quantum system, noise arises through a combination of random fluctuations in the probe and back-action of the probe on the central system \cite{caves1981quantum}. For example, in optomechanics, the system is a mechanical element and the probe is an optical field mode. The random probe fluctuations induce shot noise in the readout, and back-action of the probe comes from random radiation pressure of the input light exerting a random force on the mechanics. The total quantum measurement-added noise is the sum of these two noise contributions.

A standard benchmark level for this noise is known as the ``standard quantum limit'' (SQL). The SQL is achieved by optimizing the shot noise (which decreases with increasing input laser power) and back-action (which increases with increasing laser power) to find a total minimum. In the case of a rapid impulse delivered to a mechanical sensor of mass $m$ and mechanical frequency $\omega$, a detailed analysis gives the benchmark value \cite{clerk2004quantum}
\be
\Delta I_{\rm SQL}^2 = \hbar m\omega.
\ee
This universal formula is easy to interpret: it is the size of the ground-state momentum fluctuations of the oscillator. Numerous devices exist which operate at or even below this noise level, as we discuss in detail shortly.

In general, even in the absence of any other technical noise, one has to deal with both thermal noise and measurement-added noise. We can compare SQL-level measurement-added noise to the thermal noise estimated in equation \eqref{measnoise}. One has
\be
\frac{\Delta I^2_{\rm SQL}}{\Delta I^2_{\rm T}} = \begin{cases} \frac{\hbar Q}{4 k T \tau} & {\rm suspended} \\ \frac{\hbar m^{1/3} \rho_{\rm s}^{2/3} \omega}{P (m_a k T)^{1/2} \tau}  & {\rm gas-limited} \end{cases}
\ee
Using again mm, mg-scale detectors and high vacuum, dilution-refrigeration environments as above (see \eqref{SNRfinal-mech}, \eqref{SNRfinal-falling}), we find that the measurement-added noise at the level of the SQL is substantially larger than the thermal noise (which, in such an environment, is miniscule). This means that to achieve thermally-limited detection, one needs to go below the SQL. Numerically, with the same detector parameters, we would need $10 \log_{10} \Delta I_{\rm SQL}/\Delta I_{\rm T} \approx 35, 45~ {\rm dB}$ reduction in the measurement noise for the case of a suspended or free-falling detector, respectively. This is a fundamental problem for achieving our desired sensitivities. Fortunately, there are known ways to lower the measurement-added noise to levels below the SQL.

One option is ``squeeze'' the quantum state of the readout light \cite{caves1981quantum,purdy2013strong,clark2017sideband}. Without any squeezing, the shot noise in the probe light is limited by the light's vacuum fluctuations. By putting the light in a squeezed state, the variance in one of its canonical variables is reduced (at the expense of an increase in the conjugate variable, as required by Heisenberg uncertainty). Performing measurements with this squeezed degree of freedom can thus enable measurements below the SQL. This technique is now used in many applications, including gravitational wave detection \cite{aasi2013enhanced} and searches for axion dark matter with microwave cavities \cite{zheng2016accelerating,zhong2018results}. In practice, squeezing has so far been limited to about 20 dB, typically due to optical losses.

Another method to reduce measurement noise, which can be used in tandem with squeezing, is a back-action evading or ``quantum non-demolition'' measurement \cite{braginsky1980quantum,BRAGINSKY1990251,clerk2008back,hertzberg2010back,khosla2017quantum,danilishin2018new}. Here, instead of modifying the state of the probe light, we choose to couple it to an operator of the mechanical system which enables noise reduction. In standard optomechanical sensing, the optical field is coupled to the position variable of the oscillator. However, one could instead try to couple to the \emph{momentum} variable. For a sufficiently fast signal (such as the rapid impulse from a passing dark matter particle), the slow mechanical sensor is essentially a free particle over the course of a given event, and so its Hamiltonian commutes with $p$. Thus the measurement adds no noise to a subsequent measurement, i.e. the measurement is ``non-demolition''. 

\begin{figure}
\includegraphics[scale=0.8]{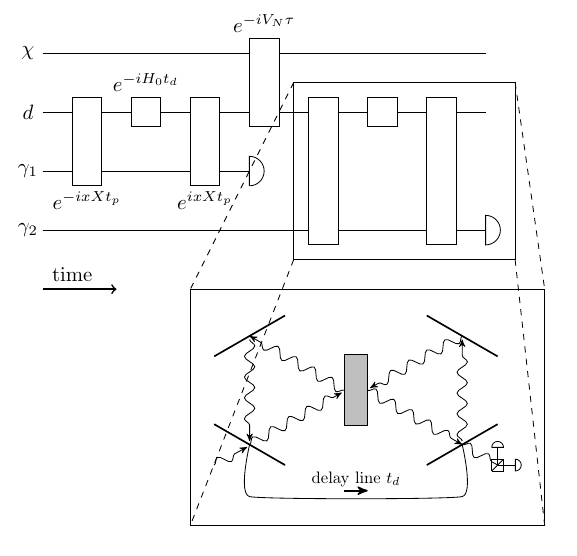}
\caption{Top: Circuit diagram depicting a backaction-evading velocity measurement. A pulse $\gamma_1$ imprints the mechanical position $x$ onto the light (described by its amplitude and phase quadratures $X,Y$). This is done twice, with opposite phase and a time delay $t_d$, leading to a velocity measurement. A second pulse $\gamma_2$ then enables a measurement of the impulse $\Delta I$. Inset: Concrete realization of a single velocity measurement, using a pair of optical ring cavities with a suspended mirror as the detector \cite{BRAGINSKY1990251,danilishin2018new,Ghosh:2019rsc}. The output light is read out via interferometry. Since the photon imprints a momentum $+p$ in the first cavity and $-p$ in the second, there is no net forcing of the mechanics: the measurement produces no quantum backaction.}
\label{velocitycartoon}
\end{figure}

It was realized long ago that back-action evasion could be used to reduce measurement noise below the SQL in a mechanical system \cite{braginsky1980quantum}. See figure \ref{velocitycartoon} and \cite{Ghosh:2019rsc} for a detailed momentum sensing protocol which in principle should exhibit around 30 dB of noise reduction below the SQL with the sensor parameters and signal considered in this paper. Experimental demonstrations of backaction-evasion exist (eg. \cite{clerk2008back}), and LIGO-scale prototypes are under current development \cite{danilishin2018new}. The noise reductions achieved have so far been modest, and again are typically limited by optical loss. However, utilization of this technique is substantially unexplored, particularly in the sub-kg scale devices considered here, so we hope that our proposal can provide impetus for new developments. 

To summarize: the fundamental technical noise floor in a high-precision mechanical sensing protocol is set by the irreducible coupling of the sensors to their thermal environments. However, continuous quantum measurements also induce a sizable source of noise. With the scale of devices considered in this paper, substantial reduction of the measurement-added noise will be required, at least a few orders of magnitude beyond what has currently been demonstrated. This presents the key challenge to realizing our proposal (besides the large number of devices). Although difficult, there is no reason in principle to believe that these noise levels cannot be achieved.

\subsection{Correlated noise between sensors}
\label{section-correlations}

So far, we have assumed that the noise on a given sensor is independent of the noise on other sensors, i.e., there are no sensor-sensor correlations. More precisely, we assumed the uncorrelated noise dominates over correlated noise. Here we make some simple estimates justifying this assumption.

Given the macroscopic nature of our sensors and small spacings, the most important effect to worry about is coming from electromagnetic potentials which couple the sensors to each other. The most important such potentials come from ``patch potentials'' (surface imperfections which carry charge) and van der Waals/Casimir forces. In fact, this is the key reason we chose our spacings to be at least $d = 1~{\rm mm}$: it is well known from eg. torsion balance experiments that these forces are sufficiently weak around this scale that they can be safely ignored \cite{PhysRevLett.103.060401}. This is the basic reason that searches for fifth-forces or modified Newtonian potentials tend to be sensitive only down to around the $50~{\rm \mu m}$ scale \cite{Kapner:2006si}.

Given that we can neglect these electromagnetic correlations, another question arises: what about sensor-sensor interactions \emph{from gravity}? Indeed, by design, our sensors are capable of measuring the sensor-sensor gravitational interactions. The question then is what is the nature of the \emph{noise} generated by this interaction. The sensor-sensor coupling is
\begin{align}
\label{coupling}
V = \frac{G_N m_s^2}{d} \frac{x_1 x_2}{d^2} \equiv m_s \Omega^2 x_1 x_2, \ \ \ \Omega = \sqrt{\frac{G_N m_s}{d^3}} \lesssim 0.1~{\rm mHz}
\end{align}
where $x_1, x_2$ are the displacements from equilibrium of the two sensors. The bound comes from assuming a reasonable solid density. Each sensor position responds to input forces according to the usual linear-response relation $x_i(\nu) = \chi_m(\nu) F(\nu)$, where $\chi_m(\nu) = \left[ m_s \left( (\nu - \omega_s)^2 - i \gamma \omega_s \right) \right]^{-1}$ is the mechanical response function for a damped harmonic oscillator. 

The sensor-sensor coupling \eqref{coupling} allows for correlated noise, namely, the input noises on sensor 2 can be transmitted to sensor 1 and vice versa. The question then is how the scale of this correlated noise compares to the uncorrelated noise. We write $F_i(\nu), i = 1,2$ for the noise acting separately on each sensor. Then sensor 1 for example has two noise terms
\be
x_1(\nu) = \chi_m(\nu) F_1(\nu) + \chi_m(\nu) m_s \Omega^2 \chi_m(\nu) F_2(\nu),
\ee
where the second term is due to \eqref{coupling}. But then for high-frequency signals like the fast impulses we are concerned with, we see that the correlated second term is suppressed relative to the first by a factor $\Omega^2/\nu^2$ due to the high-frequency behavior of the additional response function. For our gravitational problem with signals $\nu \sim 1~{\rm MHz}$, this factor is on thus of order $(10^{-4}/10^{6})^2 = 10^{-20}$, and utterly neglible compared to the uncorrelated noise terms.

The above considerations stem from the interactions between the mechanical elements in the sensors. In a practical realization, there can also be coupling through, for example, the support structure which may connect the sensors. Although important to understand, a serious study would require a detailed implementation. In particular, one may be concerned about collective modes in the array which could mimic the track signal considered above. Standard phononic engineering techniques (see eg. \cite{ghadimi2018elastic}) should be capable of controlling and/or mitigating these effects. Since this goes beyond the simple approximations of a uniformly spaced lattice as we are using here, we postpone a detailed study for further work.

In addition to the above concerns, we note also the existence of environmental noise sources which have characteristic wavelength long enough to affect multiple sensors in the array. Given millimeter or larger spacing, the dominant source of such noise would be seismic noise and gravity gradients (see eg. \cite{hughes1998seismic}). These types of noise would necessarily be at low (sub-10 Hz) frequencies, and thus do not contribute to the range of frequencies of the DM signal ($\sim 1/\tau \gtrsim {\rm MHz}$).

Finally, we note that correlations between sensors could be purposefully engineered, as a mechanism for \emph{enhancing} the sensitivity of the total detector. While relatively unexplored, this idea has been suggested as a promising route to a number of sensing goals, see eg. \cite{heinrich2011collective,chang2011slowing,xuereb2012strong}.

\subsection{Data processing}
\label{section-data}

As described above, the detector concept involves continuously monitoring a large number of devices and then looking for tracks in the data. Brute force implementation of this would be a computationally intensive process. Although looking for tracks with a billion sensors may seem daunting, it is worth noting that the next generation CMS detector at the LHC has of order two billion pixels, and will be used to search for substantially more complex track signals than the simple straight-line trajectories considered here. Understanding the computational requirements and efficient implementation of the algorithms, including compression, is a subject of current work by us. Here we make a few preliminary remarks.

Data filtering is a key component of our proposal. We are interested in searching for a signal of known temporal shape in a noisy time-series. This problem is essentially the same as faced by LIGO in the search for gravitational waves, and we suggest borrowing a technique from them (and many other signal-processing applications) known as matched filtering \cite{owen1999matched}. Here one takes the output data $F_i(t)$ with $i=1,\ldots,N$ labeling the individual sensors and convolves the data $O(t_e) = \sum_i \int F_i(t-t_e) f_i(t) dt$ with a \emph{filter} designed so that the convolution peaks on the tracks we are looking for. Here $t_e$ means the time of an event, which must be scanned over. In our problem, $f(t)$ would roughly match the time-dependent force \eqref{signal} we are looking for; as described above, we could for example use such a filter to look for any particular component of the force, not just the component perpendicular to the DM track \cite{Ghosh:2019rsc}. The effects of compression on the time-series data are controllable and should present a minor fractional change in the sensitivity \cite{herman2009high,davenport2006compressive,panelli2020applying}, although a detailed analysis of this issue in the context studied here will be an important piece of future work. A particularly appealing method could be to use discrete pulse-based measurements instead of a continuous measurement scheme \cite{vanner2011pulsed}.

The convolution over multiple sensors represents the major computational challenge in our problem. Implementing it efficiently is an important challenge which we are currently studying. We note here one key option to reduce the overhead could be to use Radon-transformed data \footnote{This idea is due to Juehang Qin.}. The Radon transformation maps functions $f:R^3 \to R$ on the detector array to the integral of the function $Rf(\gamma) = \int_{\gamma} d\lambda f(\lambda)$ along a given track $\gamma$ through the detector. Since our fundamental observable is really the impulse delivered to a track worth of sensors, it is natural to look in Radon space. This substantially reduces the computational cost because instead of looking at $N^3$ sensors one needs only to scan over the set of $\sim N^2$ independent tracks. The Radon transformation, including matched filtering, could in principle be implemented in hardware (see e.g.\ \cite{steier1986optical,ilovitsh2014optical}), which would reduce the complete computational cost to a maximization scan over $N^2$ pixels of Radon space. We are currently studying a prototype of this idea, the details of which will appear in a separate publication.

\subsection{Architectural choices}

Mechanical sensing devices operated in the classical or quantum regime come in a wide variety of architectures. These include torsion balances \cite{adams1991use,gillies1993torsion,bantel2000high}, suspended mirrors \cite{TheLIGOScientific:2014jea}, stretched membranes \cite{hertzberg2010back}, levitated dielectrics \cite{yin2013optomechanics,hempston2017force} or superconductors \cite{chan1987superconducting,chan1987superconducting,prat2017ultrasensitive}, liquid helium \cite{shkarin2019quantum}, and more. The choice of which specific type of device to use for an experiment like the one described here is well beyond the scope of this paper, but here we make a few remarks about basic issues to be considered.

The most familiar example of an optomechanical force sensor consists of a suspended mirror monitored by light, as in LIGO \cite{TheLIGOScientific:2014jea}. One may be concerned about the use of an enormous number of lasers in a small volume; to eliminate the need for these, one can instead use electromechanical couplings or fiber-coupled devices \cite{blencowe2004quantum}. An alternative approach would be to use a single laser to interrogate multiple devices; this could potentially be used to coherently read out the system.

Support structures will thermally load on the sensors; to eliminate this, one could periodically drop the detectors and allow them to freely fall for some short time, so that thermal noise comes only from collisions with ambient gas. The essential duty cycle is schematically depicted in the circuit diagram of figure \ref{velocitycartoon}. In the first step, we turn off the trap \cite{henderson2009experimental} and allow the sensor to fall. We then measure the momentum of the sensor, wait for a time of order the DM flyby time $\tau$ to let the potential DM interact with the sensor, and perform a second momentum measurement, yielding the change in momentum $I = \Delta p$. Given the short times of interest, the sensor will fall an essentially negligible amount, so this cycle can be repeated.

\section{Outlook and conclusions}

We have presented a radically new DM direct detection strategy involving a meter-scale array of high precision force sensors. Unlike traditional searches for dark matter, our technique requires no ad hoc assumptions about DM beyond its required gravitational coupling to other particles. Reaching the required sensitivity presents a clear target for development of quantum impulse measurement protocols, a concept with many applications beyond those discussed here. Although significant further work will be required to realize our scheme in detail, the potential payoff--the possibility of a direct DM detection method with no reliance on non-gravitational coupling--is enormous.

\begin{acknowledgments}
We thank Vladimir Aksyuk, Joshua Batson, Zackaria Chacko, Aaron Chou, Roni Harnik, Matthew Hollister, Dan Hooper, Mark Kasevich, Noah Kurinsky, Rafael Lang, Zhen Liu, Nobuyuki Matsumoto, Sam McDermott, Alissa Monte, David Moore, Lina Necib, Jon Pratt, Tom Purdy, Peter Shawhan, and Terry Tope for discussions, and we thank our referees for a highly instructive correspondance. Fermilab is operated
by Fermi Research Alliance, LLC, under Contract No. DEAC02-07CH11359
with the US Department of Energy. 
\end{acknowledgments}

\bibliography{GravitationalDetectionPRD-2}

\end{document}